\documentclass[12pt,a4paper]{iopart}

\usepackage{harvard}
\newcommand{\be}{\begin{equation}}
\newcommand{\ee}{\end{equation}}
\newcommand{\eqref}[1]{(\ref{#1})}
\newcommand{\text}[1]{\mathrm{#1}}

\usepackage{graphicx}

\newcommand{\citet}[1]{\citeasnoun{#1}}
\newcommand{\citep}[1]{\cite{#1}}
\newcommand{\bibspath}{/home/mleitner/aktiv/text/bibtex-referenzen/}
\newcommand{\vect}[1]{\mathbf{#1}}
\usepackage{microtype}

\begin{document}

\title{Fermi surface determination from momentum density projections}

\author{Michael Leitner$^{1,2}$, Josef-Andreas Weber$^1$ and Hubert Ceeh$^1$}

\address{$^1$ Physics Department, Technische Universit\"at M\"unchen, James-Franck-Str.\ 1, 85748 Garching, Germany}
\address{$^2$ Heinz Maier-Leibnitz-Zentrum, Technische Universit\"at M\"unchen, Lichtenbergstr.\ 1, 85748 Garching, Germany}
\ead{michael.leitner@frm2.tum.de}
\begin{abstract}
The problem of determining a metal's Fermi surface from measurements of projections of the electron or electron/positron momentum densities, such as obtained by Compton Scattering or Angular Correlation of Positron Annihilation Radiation, respectively, is considered in a Bayesian formulation. A consistent approach is presented and its advantages compared to previous practice is discussed. A validation of the proposed method on simulated data shows its systematic accuracy to be very satisfactory and its statistical precision on modest experimental data to be surprisingly good.
\end{abstract}

\maketitle

\tableofcontents

\section{Introduction}
The classical approach for experimentally determining the Fermi surface of metallic systems is to exploit its effect on quantum oscillations, such as in the de Haas-van Alphen- or the Shubnikov-de Haas-effects. These techniques, while proven to be very powerful for obtaining quantitative information on the dimensions of the Fermi surface with high precision, vitally depend on long scattering lengths of the electrons, and therefore are applicable only at cryogenic temperatures and at vanishing occupational disorder. Apart from that, the task of assigning the measured extremal orbits to actual features of the Fermi surface becomes challenging for systems with more complicated multi-sheet Fermi surfaces \citeaffixed{brasseprb2013}{e.g.,}.

Experimental methods that measure electron momentum densities work equally well for disordered states (both occupational disorder and temperature) and therefore hold the promise of determining the Fermi surface from the discontinuities in the occupied momentum densities as long as the concept of a well-defined Fermi surface is meaningful at all \citep{dugdalelowtempphys2014}. These methods comprise Compton scattering \citep{cooperrepprogphys1985} and Angular Correlation of Positron Annihilation Radiation \citeaffixed{bissonhelvphysacta1982}{ACPAR, also ACAR,}. However, in these techniques the primary experimental data are plane projections (Compton scattering, early positron annihilation) or line projections (recent positron annihilation setups), from which the three-dimensional momentum density has to be computationally determined. Also, in positron annihilation experiments the sampled two-photon momentum density differs from the underlying ideal electron momentum density due to positron wave-function effects and electron-positron correlations, although the position of the discontinuity due to the Fermi surface will remain unaffected.

The problem of reconstructing densities from projections has seen much attention due to its relevance for medical and technological imaging. For used approaches in the specific case of momentum densities in solids see, e.g., the recent review by \citet{kontrymsznajdlowtempphys2009}. In short, densities were reconstructed in the majority of previous works either by discretizing an analytical inverse of the Radon transform \citep{radonintegralwerte1917}, or by expanding the measured projections and sought densities into basis functions with convenient transformation behaviour. Both of these approaches are direct methods in the sense that the results are derived by applying a sequence of explicitly defined transformations to the data. As a consequence, their computational complexity is modest, which historically was a reason for their adoption, and for diagnostical applications still is. On the other hand, experimental methods for determining electron momentum densities are countrate-limited, therefore it should be the power of an analysis method rather than the runtime which dictates the method to be preferred.

In this paper, we will give a Bayesian formulation of the data analysis problem. For illustration, we will concentrate on the case of two-dimensional ACPAR, although our approach is equally applicable to Compton scattering. We will show that the formulation corresponds to a regularized inverse problem, and we will illustrate how its solution can be practicably obtained. The main features of our method lie in the avoidance of systematic errors thanks to a consistent description of the whole problem, and quantitative results due to an explicit parametrization of the Fermi surface. We will demonstrate the power of our method by applying it to simulated data, which will allow us to conclude that an ACPAR experiment with moderate statistics and resolution is able to determine the shape of the Fermi surface \emph{quantitatively} with an accuracy that is comparable to quantum-oscillatory methods, at elevated temperatures and in the presence of disorder.

\section{Definition of the problem}
\subsection{Bayesian formulation}
In the problem at hand, the basic unknown quantity is the three-dimensional momentum density $\rho(\vect{p})$. For Compton scattering, this concerns the actual electron momentum density (Fourier components of the electron states), while in ACPAR the positron wave-function and electron-positron correlation effects modify the probed density, which is then termed two-photon momentum density. In either case, the density is a smooth function, apart from steps when crossing sheets of the Fermi surface. In the extended-zone scheme, the density will decay towards high $\vect{p}$ and display the point group symmetry of the crystal, while the Fermi surface features have the crystal's space group periodicity.

The experimental data $y$ are given by a set of one- or two-dimensional spectra, corresponding to plane or line projections of the underlying three-dimensional densities. To a very good approximation, the data follow Poissonian statistics, in particular the noise for distinct data points is independent, and the noise probability distribution is uniquely defined by the expectation (and can be estimated from the measured signal). 

The relation between these two quantities is given by a linear operator $\mathbf{P}$, which specifies how the expected value of the measured signal (i.e., before Poisson quantization) results from a given, but experimentally a priori unknown, momentum density. We will call it the projection operator, as its function is to essentially integrate over the transversal or longitudinal momentum components for the distinct experimental orientations of the crystal. Additionally, it also takes the detector efficiency and momentum sampling functions and the smearing of the spectra due to finite resolution into account, which will be discussed in more detail in Sect.\ \ref{operators}.

Thus, our initial formulation for the problem to be solved is the following: What is the posterior probability distribution $p_\text{post}$ for $\rho$, given an outcome of the experiment $y$ and considering prior knowledge or assumptions, quantified in the prior distribution $p_\text{prior}(\rho)$? By Bayes' formula, the answer is formally given by
\be
p_\text{post}(\rho|y)=\frac{p_\text{lik}(y|\rho)p_\text{prior}(\rho)}{p(y)},\label{bayes1}
\ee
where $p_\text{lik}(y|\rho)$ is the so-called likelihood function, the modelled probability for observing the actual experimental outcome $y$ in a repeated experiment under the assumption of the momentum density $\rho$. In the present case it is just
\be
p_\text{lik}(y|\rho)=\prod_ip_\text{Poisson}\Bigl(y_i,\bigl(\mathbf{P}(\rho)\bigr)_i\Bigr)=:P_\text{Poisson}\bigl(y,\mathbf{P}(\rho)\bigr),
\ee
where the measured spectra are treated as a vector $(y)_i$ and 
\be
p_\text{Poisson}(k,\lambda)=\frac{\lambda^k}{k!}\mathrm{e}^{-\lambda},
\ee
is the familiar expression for the Poisson distribution, with $P_\text{Poisson}$ its version for vector arguments.

The challenge to the physicist lies in formulating an expression for $p_\text{prior}(\rho)$ that takes into account the available understanding of the problem and therefore ensures a physically meaningful result. A first step towards this goal is the strict requirement for $\rho$ to conform to the point symmetry group of the crystal. Instead of encoding this via a $\delta$-like part in $p_\text{prior}(\rho)$, it is more efficient to describe the densities only by their values in the irreducible wedge according to the point symmetry, with appropriate continuation over all of the momentum space. From here on, $\rho$ is to be understood in this sense.

The defining idea of our proposed approach actually follows just from letting modelling be guided by the physical picture: The expected behaviour of the density (smooth variations apart from jumps at the Fermi surface) is due to its being composed of contributions from the distinct conduction bands, each multiplied by a Fermi-Dirac occupation function with a practically discontinuous jump, sitting atop the contribution from the core states. The bare band densities (i.e., before taking occupation into account) will in fact be smooth, as can be derived from any simply tractable band structure model (be it nearly free electrons, tight binding, or the effective potentials in density functional theory). Instead of considering the total occupied density displaying jumps at the Fermi surface as the fundamental unknown, the most natural formulation therefore is to have as free parameters both (1) a parametrization of the Fermi surface sheets in each conduction band (and spin channel in the case of magnetic ordering) and (2) the bare densities of the respective conduction bands plus the summed contribution from the core bands. The prior distribution $p_\text{prior}(\rho)$ can then be used to favour smooth band densities with, e.g., additional positivity constraints or assumptions on the decay with $\vect{p}$, or also information on the Fermi surface shapes, if available from prior experiments or calculations. Note that also for choosing the parametrization of the Fermi surfaces the physical picture indicates the natural way as the level set of a smooth function with the space group symmetry (in other words, the band dispersions).

With these observations, we can rewrite Eq.~\eqref{bayes1} as
\be
p_\text{post}(\rho,\sigma|y)\propto P_\text{Poisson}\Bigl(y|\mathbf{A}_\sigma(\rho)\Bigr)p_\text{prior}(\rho,\sigma)\quad\text{with}\quad\mathbf{A}_\sigma=\mathbf{P}\mathbf{X}_\sigma,\label{bayes2}
\ee
where $\rho$ is now to be understood as the bare band densities and the action of $\mathbf{X}_\sigma$ for given Fermi surface parameters $\sigma$ is to multiply the bare band densities by the occupations and sum over the band index, i.e., it is again a linear operator.

If the primary quantity to be determined is the Fermi surface, Eq.~\eqref{bayes2} can be marginalized over $\rho$ to give the posterior distribution of the Fermi surface parameters
\be
p_\text{post}(\sigma|y)=\int\mathrm{d}\rho \,p_\text{post}(\rho,\sigma|y).\label{marginal}
\ee
Note that as we will elaborate below, the width of $p_\text{post}(\rho,\sigma|y)$ with respect to $\rho$ does not vary much as a function of $\sigma$, so for practical purposes $p_\text{post}(\sigma|y)$ is proportional to the maximum of $p_\text{post}(\rho,\sigma|y)$ for given $\sigma$. 

\subsection{Comparison to previous approaches}
The main points that distinguish our approach from the majority of those used previously are the following: first, it is formulated as a general problem of Bayesian inference instead of as an recipe of transforms obtained from analytical manipulations, and second, the Fermi surface is treated explicitly during the reconstruction instead of determined afterwards from the reconstructed densities. Here we will discuss the implications of these differences.

The Radon transform is a bijection (in particular, it is invertible) between suitably regular $n$-dimensional function spaces \citep{radonintegralwerte1917}. Specifically, it corresponds to the relation between a pointwise defined function and its integrals over all lines in the plane (in two dimensions), or its integrals over all planes in space (in three dimensions). This has the consequence that in two-dimensional ACPAR, where the accessible data would in principle be the integrals over all lines in space (a four-dimensional manifold), typically only projections within a certain plane of rotation are considered, which reduces the three-dimensional problem to independent two-dimensional problems that can be solved by direct algorithms for the two-dimensional inverse Radon transform. While in the case of medical imaging such a sectioning approach is appropriate for minimizing the necessary radiation doses due to the elongated shape of the human body, this does not apply for projections of momentum densities in reciprocal space and thereby corresponds to neglecting potentially independent information. In fact, due to the point group symmetry of the crystal taking only projections within a common rotation plane already implies information about out-of-plane projections in many cases, which a pure sectioning approach does not take into account. In a recent proposal \citep{kontrymsznajdapplphysa2007} this deficiency is addressed by re-parametrizing the reconstructed slices in terms of basis functions with the appropriate symmetry, but it is not known to which extent this can recover the information that was initially available, considering the correct symmetry, or whether the filtered reconstruction still reproduces the measured projections. In addition, the direct approaches typically need evenly and densely spaced projections in the plane of equal statistical precision, where the angular increment corresponds to the smallest resolvable features. In contrast, our proposed formulation allows an arbitrary number of projections with arbitrary orientation and noise level to be used, and takes into account all available information considering the prescribed symmetry. An early suggestion by \citet{pecoraieeetransnuclsci1987} to solve directly for the coefficients of three-dimensional basis functions of appropriate symmetry, dealing with the above-mentioned dimensional problem in a least-squares sense, should in principle be of equivalent quality in terms of these criteria, but apparently has seen only limited use. Note that it was shown there that it is not merely a welcome option to be able to use arbitrary orientations, but that using low-symmetry projections is actually indicated for optimal results in reconstruction.

Direct transform methods are derived from analytical inversions of the mathematical Radon transform. As such, they cannot cope with experimental subtleties such as finite resolutions. Therefore, the resolution is typically deconvolved from the measured spectra before reconstruction. In most cases, maximum-entropy regularization is used to solve this ill-posed problem, either explicitly \citep{fretwellepl1995} or implicitly \citep{gerhardtprb1998}, and sometimes even more arbitrary methods are used \citep{chibaphysstatsolc2007}. In contrast, aggregating all experimental complications (e.g., resolution, momentum sampling, shifting) into the forward operator as proposed here allows us to solve the inverse problem in a single step, subject to prior assumptions (equivalent to regularization) on the fundamental physical quantities, and thereby to rule out a compounding of the potentially conflicting regularization biases introduced at sequential steps. A very welcome additional consequence of our not touching the experimental spectra at all is that the noise statistics of the spectra remain unadulterated, specifically the Poissonian behaviour with independence between neighbouring pixels is conserved, which allows us to propagate the uncertainty of the data consistently into an estimated error of the final reported quantities.

The effect of experimental noise on the reconstructed densities is a critical issue for the conventional methods. Especially with the direct transform methods that rely on the central slice theorem and interpolation \citep{hondophysstatsolb1993}, the reconstruction is typically very ill-defined around the origin due to the conflicting information, and there is no obvious way of countering this problem. Such an effect is present also for basis expansion methods \citep{pecoraieeetransnuclsci1987}. In general, the regularity of the reconstructions is controlled by the number of considered basis functions in the case of expansion methods, and by the use of an appropriate filter function in the various filtered transform methods, which in either case is rather opaque to the user. In contrast, in our formulation regularization by way of the prior distribution is explicit, and no noise artefacts appear in the reconstruction. Note that an explicit regularization functional is also the only practicable way to have non-linear biases such as a non-negativity constraint \citeaffixed{pylakapplphysa2011}{as already observed by}. 

In most published works, the identification of the Fermi surface is done subsequent to and independent from the reconstruction of the density. The simplest option is to transform the density from $\vect{p}$- to $\vect{k}$-space, i.e., subject it to the so-called LCW folding \citep{lockjphysf1973}, and define the Fermi surface as an iso-density contour, e.g., according to a maximum gradient criterion \citep{biasiniprb2002}. The problem with such an approach in positron annihilation experiments is that due to the inequivalence between the measured two-photon momentum density and the electron momentum density, filled bands give rise to a non-constant background \citep{lockapplphys1975}, so that for finite resolution the Fermi surface does not strictly correspond to any iso-density contour. Edge-detection or enhancement methods \citeaffixed{dugdalejphyscondmat1994,obrienjphyscondmat1995}{e.g.,} should be able to obviate this issue, but also in this case finite experimental resolution will tend to smooth regions of the Fermi surface with high curvature. Due to the explicit treatment of the Fermi surface during reconstruction, our proposed method does not suffer from the above-mentioned effects. We are aware of only two comparable proposals: \citet{biasiniphysb2000} determined the parameters of a model Fermi surface by minimizing the deviation from the measured spectra after LCW folding, which is essentially the same idea as our proposal, only with constant band densities in $\vect{k}$-space. This idea of reconstruction by a piecewise-constant function with explicit treatment of the step manifold has also been suggested in the context of diagnostical imaging \citep{ramlaujcomputphys2007}. Second, \citet{laverockprb2010} propose to fit the LCW-folded spectra by calculated band densities, where the free parameters correspond to a state-dependent annihilation enhancement and energy shifts of the rigid bands. Clearly, in such an approach the freedom in the shape of the Fermi surface is very restricted, and the results will depend on the correctness of the electronic structure used as input.

The only disadvantages of our proposed approach that we are aware of concern the increased numerical effort compared to direct methods, although this has ceased to be relevant with today's computing power as we will show below. Also, no packaged software exists yet, but actually for momentum density reconstructions, different from medical imaging, typically custom implementations are used in any case, which especially in the case of orthogonal expansions can become more involved than our implementation as sketched below.

\section{Implementation}
Due to the large dimensionality of the problem defined by \eqref{bayes2}, an efficient way to arrive at its solution is imperative for its practicability. We will give a detailed discussion of our implementation below, for its numerical aspects see the Supplementary Material.

\subsection{Parametrization of the solution space}
In our formulation, both the band densities $\rho$ and the Fermi surface sheets $\sigma$ are explicit degrees of freedom. As mentioned above, the physical picture suggests to describe the Fermi surface as the level sets of auxiliary smooth functions with the appropriate symmetry. For this purpose we use a Fourier description, where the required reciprocal-space translation symmetry is enforced by considering only those Fourier coefficients that correspond to real-space lattice vectors. Note that this is formally identical to a tight-binding description, and in fact it has been shown that it can reproduce the experimental Fermi surfaces in the noble metals with only a few free parameters \citep{roafphiltransa1962}. In contrast, for systems that conform rather to the free-electron picture, such as Al, different models will probably be more efficient in describing the Fermi surfaces \citep{ashcroftphilmag1963}.

Thanks to the explicit treatment of the Fermi surface, the band densities $\rho$ will be smooth and can therefore be described with comparatively low resolution. Here we use quadratic B-splines. The point symmetry is fulfilled by considering only coefficients within the irreducible wedge, with appropriate continuation over all of the reciprocal space. For reasons of efficiency we employ a non-constant sampling density, with higher resolution at low momentum. This is justified by the observation that the contribution at high momenta is mainly due to the core electrons, which will not be influenced much by the crystal structure and therefore be nearly isotropic. 

We want to emphasize that the choice of basis functions is not essential to the idea of the method. Considerable mathematical effort has been expended in deriving expansions that fulfill orthogonality relations of some sense under idealized projection operations \citep{louissiamjmathanal1984}. In principle, such parametrizations could be used also in our proposal, where we expect that a sufficient description could be attained already at lower degrees of the expansion compared to previous implementations, as the parametrization needs to capture only the variation of the band densities excluding the Fermi surface steps. As an additional advantage, the system of linear equations to be solved would be better conditioned due to the orthogonality properties. However, such basis functions are typically non-vanishing nearly everywhere in reciprocal space, corresponding to full instead of sparse projection operators, which would constitute a serious drawback as we will discuss below.

\subsection{Matrix form of the operators}\label{operators}
The projection operator $\mathbf{P}$ specifies how a given three-dimensional momentum density leads to the set of two- or one-dimensional spectra in the specified orientations by line or plane projections. For the construction of this matrix we compute for each voxel the set of pixels or bins in each spectrum that can potentially have an overlap with the projection of the voxel, and set the respective entry to the proportion received by the corresponding pixel. We find this proportion by an approach based on a look-up table with linear interpolation, pregenerated from a high-resolution projection of a single voxel cube. Given the finite experimental resolution (discussed below), this approach can be considered as equivalent to the exact solution.

The optimal choice for the discretization of the three-dimensional momentum density is somewhat finer than the experimental resolution: A coarser representation will lead to artefacts, while a finer discretization will be numerically more expensive without any effect on the spectra after the application of resolution smearing. The same reasoning holds for the discretization of the spectra. As a consequence, each voxel will contribute to only a few pixels in each spectrum, allowing us to precompute and store the projection matrix in sparse format.

By setting the resolution kernel to zero once its value has dropped below some threshold, also the resolution smearing operator has the form of a sparse matrix. The final step in obtaining the spectra from a given momentum density is the pointwise multiplication by the momentum sampling function, which can corresponds to a diagonal matrix. In principle, the projection operator $\mathbf{P}$ is then the product of these three precomputed sparse matrices, although due to the involved dimensionalities it is more efficient to utilize associativity of matrix multiplication in the further steps and never compute the product of the resolution smearing matrix and the projection matrix proper explicitly.

The occupation operator $\mathbf{X}_\sigma$, computing the momentum density from the band densities and the Fermi surface $\sigma$, is for fixed $\sigma$ represented as the horizontal concatenation of diagonal matrices, with the entries corresponding to the respective occupations. Here special care has to be taken with the voxels at the Fermi surface: For a continuous variation of the resulting spectra with the shape of the Fermi surface, the occupations have to be computed according to the proportions of the voxel within the Fermi surface (at the relevant temperatures the width of the Fermi surface compared to the voxel size is negligible). We accomplish this by the Gilat-Raubenheimer method \citep{gilatpr1966}, i.e., by linearizing the variation of the band energy within the voxel and computing the enclosed volume explicitly. 

Summing up, the linear operator $\mathbf{A}_\sigma$ that relates the band density parameter vector $\rho$ to the spectra for a given shape of the Fermi surface $\sigma$ is the matrix product $\mathbf{P}\mathbf{X}_\sigma$ of the matrices discussed above, modified by an additional degree of freedom corresponding to a constant background intensity.

\subsection{Algebraic solution of the Bayesian problem}
The fact that allows for an efficient computation of \eqref{bayes2} lies in the observation that $p_\text{lik}(\rho,\sigma|y)$ can be approximated very accurately by a Gaussian distribution for fixed $\sigma$. Specifically, by equating the measured experimental spectra with their true value (i.e., the expected value before quantization) and expanding the logarithm of the Poissonian distribution around the maximum we have
\be
\log\bigl(p_\text{lik}(\rho,\sigma|y)\bigr)=-{\textstyle\frac{1}{2}}(\mathbf{A}_\sigma\rho-y)^\top\mathbf{W}(\mathbf{A}_\sigma\rho-y)+\text{const},
\ee
where the weighting matrix $\mathbf{W}$ is the inverse of the covariance matrix. This initial problem can be iteratively refined by expanding around the updated expected values. In our tests, such an iterative approach converged rapidly (within one iteration), and even the initial distribution was a faithful representation of the final result, as the counts per pixel were not too low and the residuals were small.

To guarantee smooth reconstructed band densities, we formulate our prior distribution as a Gaussian distribution of the square norm of the band densities' second derivatives plus an analogous contribution from the first derivatives to favour monotonicity. A non-constant weighting of these norms can be used to penalize the same density curvatures more if they occur at high momenta, where the absolute values of the densities are smaller. Our choice is functionally equivalent to Tikhonov regularization \citep{tikhonovsovmathdokl1963}, i.e., a positive semidefinite quadratic form of the parameters, when searching for the maximum a posteriori estimate. 

As both $p_\text{lik}$ and $p_\text{prior}$ are multivariate Gaussian distributions, so is their product $p_\text{post}$. As a consequence, the maximum a posteriori estimate $\rho^\star(\sigma)$ for given $\sigma$ can be obtained by solving the system of linear equations
\be
(\mathbf{A}^\top_\sigma\mathbf{W}\mathbf{A}_\sigma+\sum_i\lambda_i\mathbf{D}_i^\top\mathbf{D}_i)\rho=\mathbf{A}^\top_\sigma\mathbf{W}y,\label{lineq}
\ee
where $\mathbf{D}_i$ are the matrices that compute the (optionally weighted) derivatives of the band densities from the parameters and $\lambda_i$ the corresponding regularization parameters. Also the marginal posterior distribution with respect to $\sigma$ as defined by \eqref{marginal} follows easily as
\be
p_\text{post}(\sigma|y)=p_\text{post}\bigl(\rho^\star(\sigma),\sigma|y\bigr)\sqrt{\det(\mathbf{A}^\top_\sigma\mathbf{W}\mathbf{A}_\sigma+\sum_i\lambda_i\mathbf{D}_i^\top\mathbf{D}_i)}.\label{marginalpost}
\ee


\section{Application to model data}\label{validation}
Most previous proposals for algorithms to reconstruct three-dimensional momentum densities have been validated only by comparing the results on experimental data to those of other algorithms, if at all \citeaffixed{kontrymsznajdapplphysa2008,pylakapplphysa2011}{e.g.,}. Obviously this is not satisfactory, as specific aspects of the data could have been missed simultaneously by both the tested and the benchmark methods. In other cases the algorithms have been applied to synthetic data obtained by comparatively simplistic models, but also there the comparison has been done only in qualitative terms \citeaffixed{pecoraieeetransnuclsci1987}{e.g.,}. As we claim here to be able to reconstruct Fermi surfaces quantitatively, we have to substantiate this claim by demonstrating both its fidelity (the magnitude of introduced systematic errors) and statistical performance (the propagation of experimental noise to the resulting dimensions) on realistic data for which the correct solution is known. 

\begin{figure}[t]
  \raggedleft
  \includegraphics{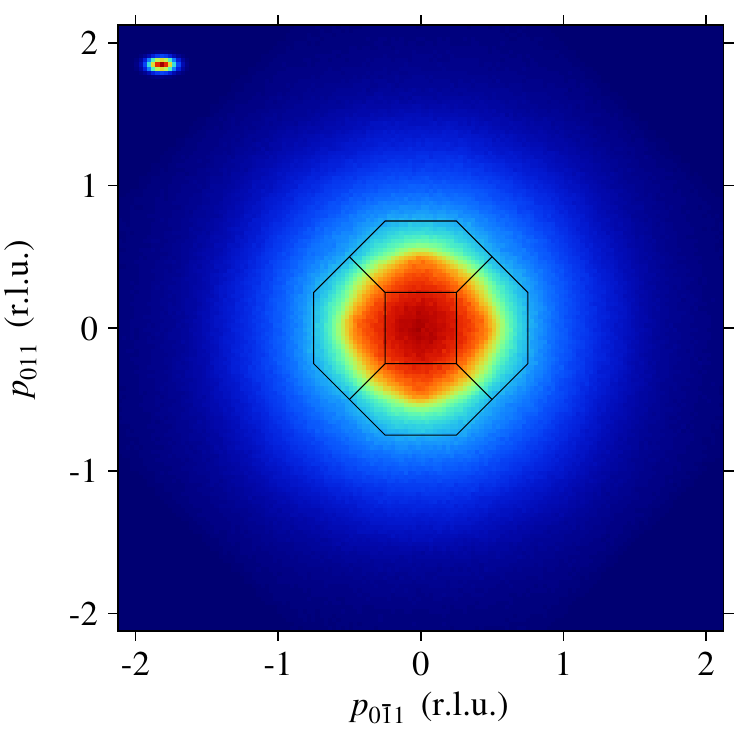}\quad\includegraphics{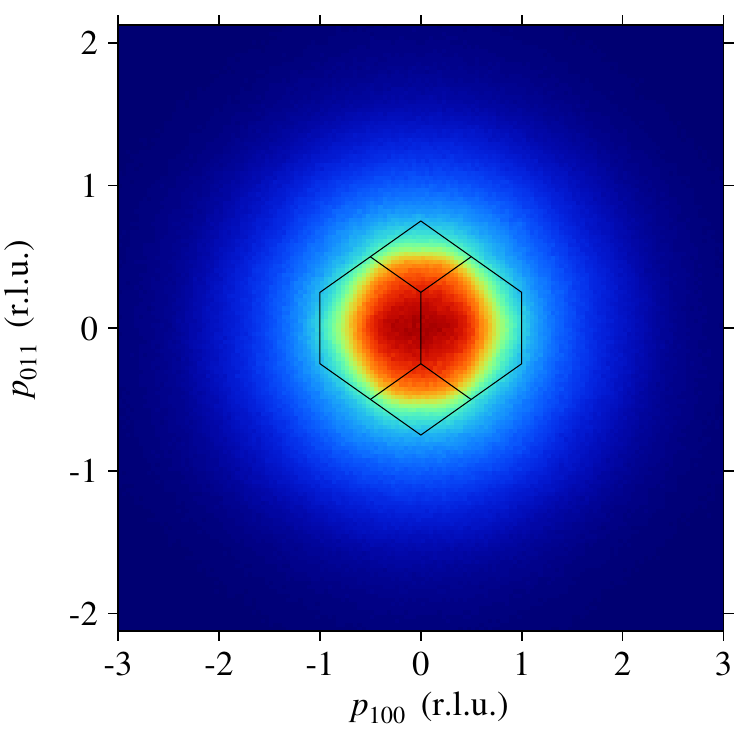}
  \caption{Simulated spectra for $(100)$ (left) and $(0\bar{1}1)$ (right) orientation, including projection of the first Brillouin zone. The assumed resolution function is indicated.}\label{projections}
\end{figure}

For this purpose we chose the system of copper with its prototypical and well-known Fermi surface. We computed the electronic structure of Cu self-consistently in the generalized gradient approximation with the PBE exchange-correlation functional \citep{perdewprl1996} by the density functional code \textsc{abinit} \citep{gonzecomputphyscomm2009}. With the converged density we computed both the electron wave functions and energies on a fine mesh and the $\Gamma$-point positron wave function \citep{barbielliniprb1995}. Thanks to the plane-wave formulation used in the \textsc{abinit} code, the electron-positron momentum densities could be conveniently derived in a custom implementation corresponding to the independent particle model. From the three-dimensional density plus some constant background we computed the corresponding ACPAR spectra of $144^2$ pixels for $(001)$, $(110)$ and $(111)$ orientation, each with $25\cdot 10^6$ counts distributed according to Poissonian statistics. We chose a discretization of 24 pixels per reciprocal lattice constant and assumed an anisotropic Gaussian resolution function with $2\times 1$ pixel standard deviation. Two of the resulting spectra are given in Fig.~\ref{projections}. With the actual lattice constant of Cu, our chosen resolution would correspond to $1.32\times 0.66\,\text{mrad}^2$ FWHM.

\begin{figure}[t]
  \includegraphics{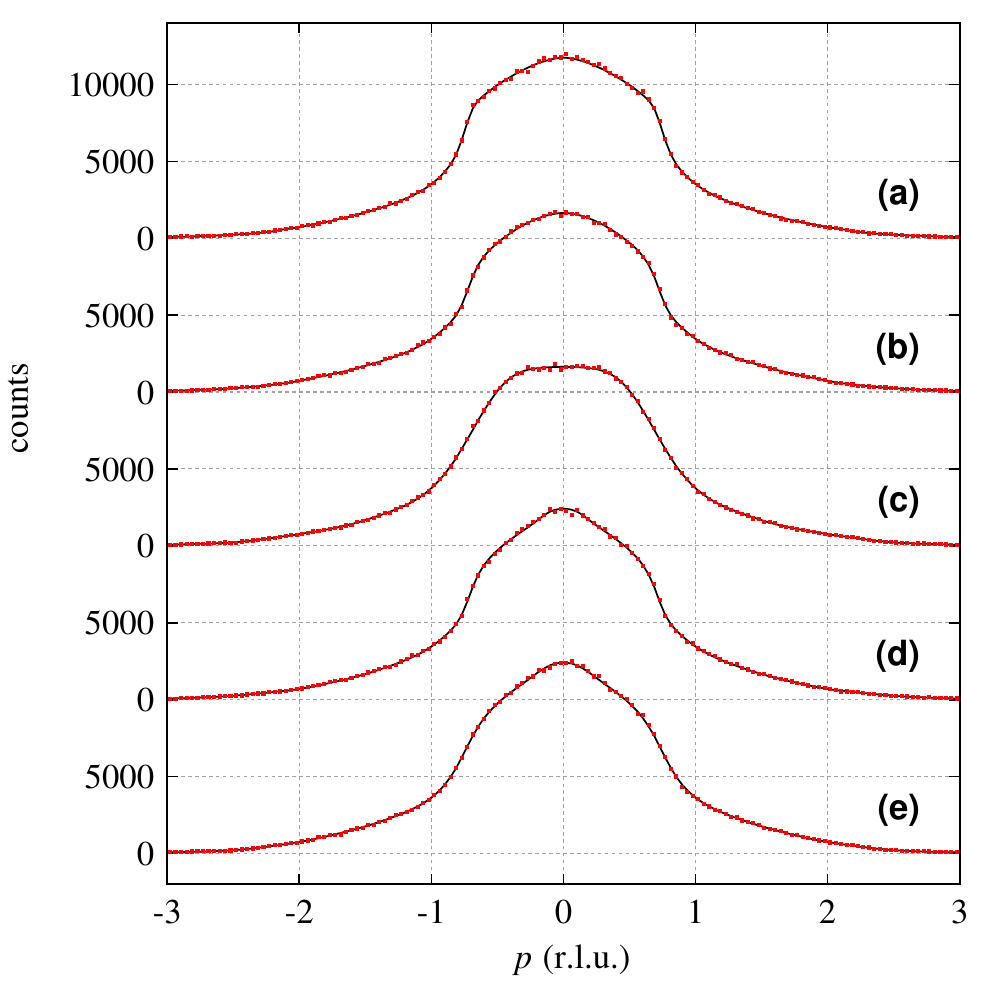}
  \caption{Selected cuts through the spectra: in $(100)$-spectrum along $(011)$ (a), in $(0\bar{1}1)$-spectrum along $(011)$ (b) and $(100)$ (c), in $(1\bar{1}1)$-spectrum along $(011)$ (d) and $(21\bar{1})$ (e), in each case through the origin.}\label{cutsspectra}
\end{figure}

We reconstructed the density in a volume of $144^3$ voxels at the discretization of 24 voxels per reciprocal lattice constant. We want to emphasize that the coincidence of pixel and voxel size is by no means necessary; specifically for an actual experiment it will be beneficial to choose the voxel size as an integer fraction of the reciprocal lattice constant, while the discretization of the spectra will be given by the apparatus. As Cu has only a single band crossing the Fermi energy, we considered a fully occupied core level electron density and a single conduction band. We parametrized the space of band densities by cubic B-splines with an increasing sampling density towards small $\vect{p}$, corresponding to 385 degrees of freedom per band, and we described the single Fermi surface sheet by a five-parameter Fourier expansion (where the $\langle 110\rangle$ coefficient is fixed to 1 as it corresponds to a trivial scaling of the energy range, and the $\langle 000\rangle$ coefficient is chosen relative to the Fermi energy so as to constrain the occupied volume to half the Brillouin zone). We also added a term to the prior distribution to favour the decay of the coefficients with interaction range, because, as already noted by \citet{roafphiltransa1962}, the Fermi surface changes only by very small amounts under certain modifications of the parameters.

Maximizing the posterior probability given by Eq.~\eqref{marginalpost} for the simulated spectra gives reconstructed spectra that, apart from the missing noise, are visually identical to the input spectra. This is substantiated by the one-dimensional cuts through the spectra shown in Fig.~\ref{cutsspectra} and the corresponding reduced $\chi^2$ value of 1.047. The degree of achieved faithfulness to the data obviously depends on the choice of the regularization parameters $\lambda_i$ in Eq.~\eqref{lineq}. Here we used the smallest values that still suppress visually noticeable artefacts (that would correspond to the reconstruction of experimental noise) in the reconstructed densities.

\begin{figure}[t]
  \raggedleft
  \includegraphics{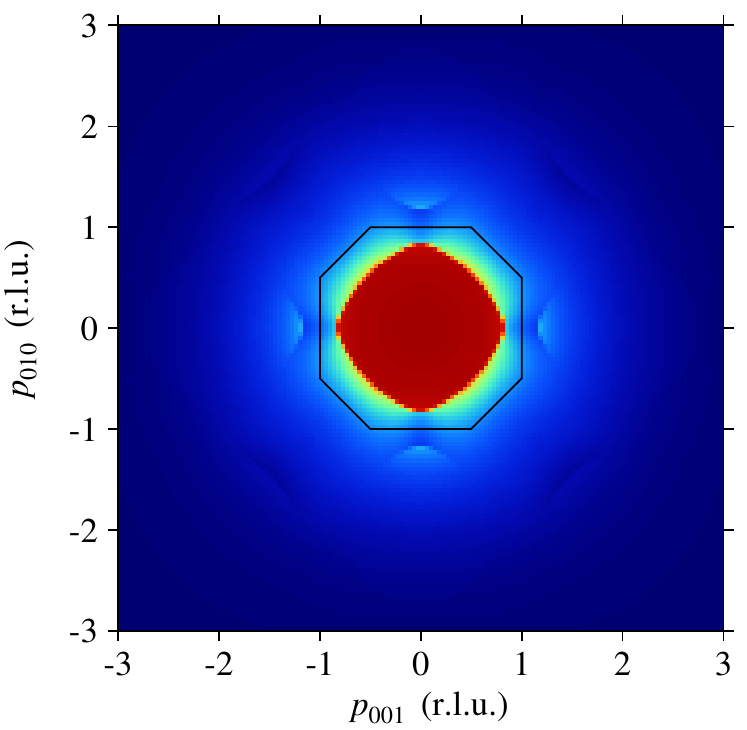}\quad\includegraphics{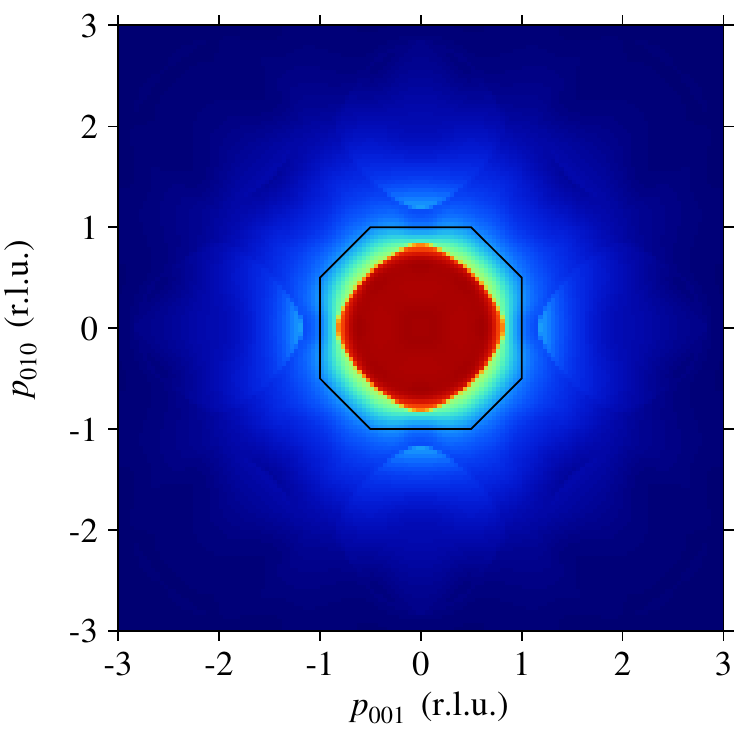}\\[.2cm]
  \includegraphics{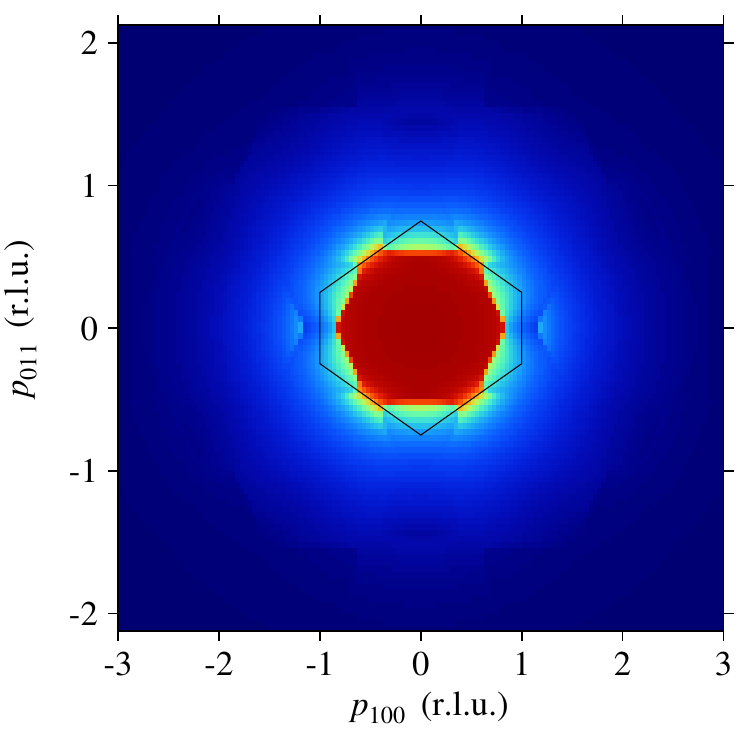}\quad\includegraphics{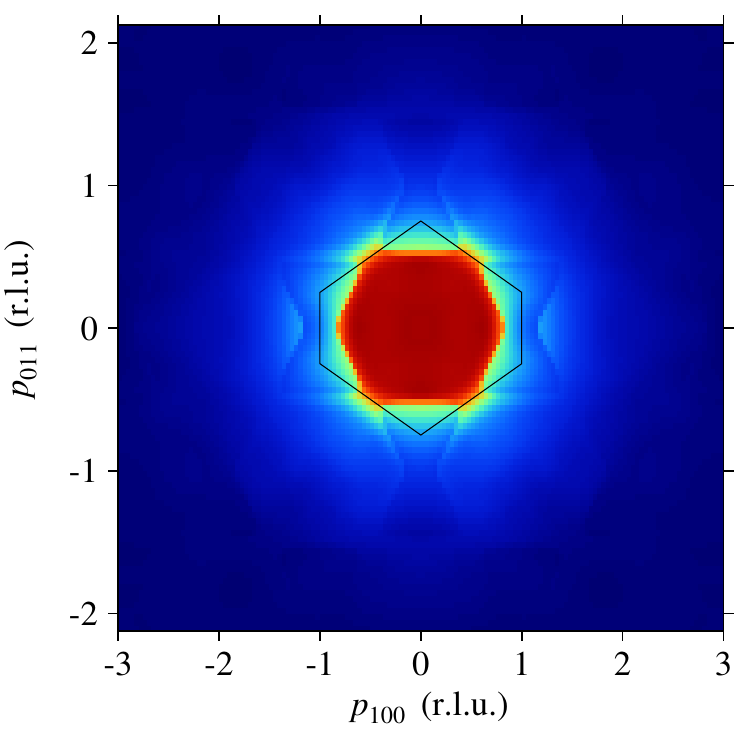}
  \caption{Cuts through the three-dimensional electron-positron momentum density. Input densities due to model (left column) and reconstructed densities (right column), $(100)$ (top row) and $(0\bar{1}1)$ (bottom row) planes through the origin, including outline of first Brillouin zone.}\label{cuts}
\end{figure}

A comparison of input and reconstructed density is given in Fig.~\ref{cuts}. Here significant systematic differences can be discerned. Specifically, the original density is virtually constant within the Fermi surface in the first Brillouin zone and decays quite abruptly through the necks into the second Brillouin zone. In contrast, the tendency of the reconstruction towards smooth variations in the band densities leads to a decrease already within the Fermi surface in the first Brillouin zone and to higher contributions from outer zones. Apart from that, the main features are clearly reconstructed in a qualitatively correct way. 

For assessing the algorithm's performance in determining the Fermi surface in quantitative terms, we focus on three specific features: the extent of the Fermi surface along (100), along (110), and the radius of the (111)-neck (its deviation from a circular shape is negligible). In the input data, these three dimensions are $1.063$, $0.946$ and $0.203$, respectively, measured in units of $r_\text{f}$, the radius of the free-electron Fermi sphere. In the reconstructions, the corresponding values are $1.063(3)$, $0.949(1)$ and $0.182(6)$, denoting the average and the standard deviations of the maximum-a-posteriori values for different realizations of the counting noise. For a single realization of the counting noise, the posterior probability distribution defined by Eq.~\eqref{marginalpost} is to a good approximation a Gaussian distribution, with a covariance matrix that essentially corresponds to above-quoted standard deviations, which would allow us to estimate the errors of the dimensions obtained from an experiment.

The small statistical uncertainty of the results indicates that our approach can also be used for qualitative statements. For instance, constraining the Fermi surface parameters in the reconstruction so that the necks along (111) become disconnected gives a maximum-a-posteriori probability that is smaller by 140 orders of magnitude, i.e., in a fictitious experiment this scenario could essentially be ruled out.

\section{Discussion}
The fact that our reconstruction reproduces the measured spectra perfectly within the errors shows that it considers all the information present in the data. Within these boundary conditions it yields the reconstruction most probable according to the prior assumptions. We think that our assumptions are arguably the soundest on physical grounds, and definitely the most transparent and easiest to adjust compared to those inherent to direct methods.

\begin{figure}[t]
  \centering
  \includegraphics[width=.45\textwidth]{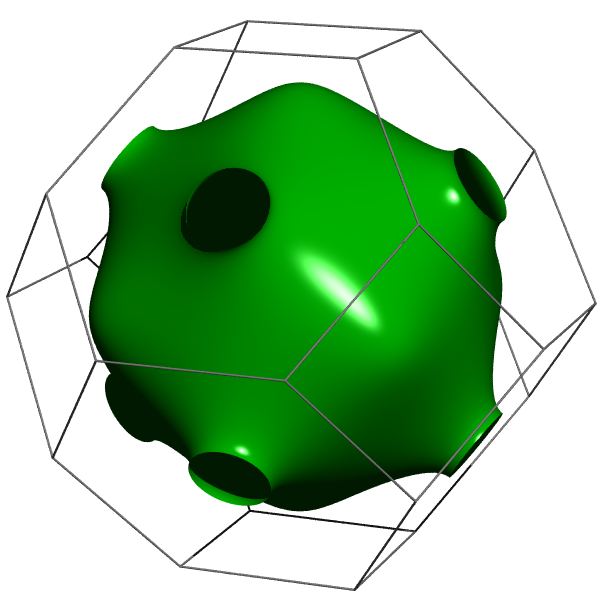}\qquad\includegraphics[width=.45\textwidth]{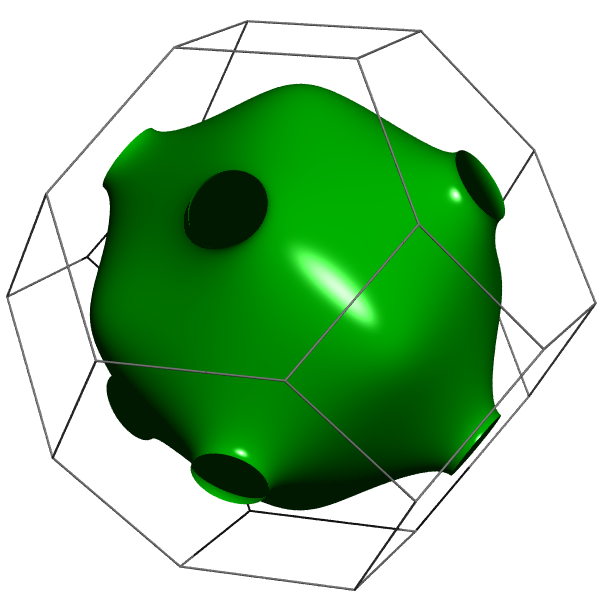}
  \caption{Rendering of the Fermi surface within first Brillouin zone, input data (left) and representative reconstruction from simulated spectra (right).}\label{fermirender}
\end{figure}

Our results show that three ACPAR spectra with moderate statistics and unexceptional experimental resolution suffice for our interpretation method to give statistically very well-defined results on the Fermi surface dimensions. In terms of accuracy it has to be noted that specifically the radius of the (111)-necks is underestimated by 10\%. However, in the overall picture this corresponds only to an error of 2\% of the mean Fermi surface radius, and all other regions of the Fermi surface are determined still much more accurate (see the juxtaposition of renderings of the input and reconstructed Fermi surface in Fig.~\ref{fermirender}). For comparison, the Fermi surfaces of Ag and Au differ much more from the actual Fermi surface of Cu than the reconstruction does with a (111)-neck of 0.137 in Ag and a (100)-radius of 1.135 in Au \citep{roafphiltransa1962}. Also, if a plausible model for the electronic structure in a given system is available, an analysis as presented here can be done on the model and the experimental results can be corrected in first approximation for the systematic effects displayed by the model reconstructions. Note further that also in quantum-oscillatory methods the Fermi surface has to be reconstructed from the measured data, subject to some plausibility assumptions.

In the interpretation of an actual experiment it would probably be too optimistic to expect a $\chi^2$ as low as reported here due to experimental imperfections such as additional contributions to the spectra, e.g., from surface positronium ejection, or slight misorientations. However, due to the adaptibility of our formulation such effects can be included at the cost of a few additional free parameters. This is in contrast to direct methods with, e.g., the strict assumption of crystal symmetry in the plane of sample rotation.

After the alkali metals, the noble metals with Cu as the example chosen here have the simplest Fermi surface and are therefore probably the easiest systems to consider. For systems with multiple Fermi surface sheets separated only by a small distance the problem will become ill-defined due to resolution effects. Due to an analogous reasoning also large real space cells and consequently small reciprocal space features make a system hard to solve. However, these limitations obviously apply equally to any method of data interpretation. 

The last point we want to stress is that in principle any additional information can be considered in the prior probability assumptions. For example, in the case at hand the ill-defined (111)-neck radius could be fixed to the value given by de Haas-van Alphen-measurements, where it corresponds to a prominent and well-defined frequency \citep{shoenbergphiltransa1962}, while the shape of the Fermi surface belly is more directly encoded in the positron annihilation data.

\section{Conclusion}
Here we have presented a new point of view on obtaining the shape of the Fermi surface from Angular Correlation of Positron Annihilation Radiation, or equivalently Compton Scattering, data. We have pointed out the advantages of a unified formulation as an inverse problem in the Bayesian setting instead of the conventional sequential approaches. We have shown that modest requirements on the data statistics lead to statistically well-defined results on the Fermi surface dimensions, and we have discussed the small systematic biases introduced by the prior assumptions for the example of copper. These insights open the way for obtaining Fermi surface information with a quality that is comparable to quantum-oscillatory methods under conditions such as high temperature and occupational disorder, where the classical methods cannot be applied.

\section*{Acknowledgments}
This work was supported by the Deutsche Forschungsgemeinschaft (DFG) through TRR 80.

\section*{References}
\bibliography{\bibspath abkuerz,\bibspath acpar,\bibspath fermisurf,\bibspath dos,\bibspath tomographie,\bibspath imaging,\bibspath abinit,\bibspath dummy}
\end{document}